\documentclass[10pt,a4paper,notitlepage]{article}
\usepackage[utf8]{inputenc}
\usepackage{amsmath,amsfonts,amssymb}
\usepackage{graphicx}
\usepackage{hyperref}

\hypersetup{
	colorlinks=true,         % false: boxed links; true: colored links, false is default
	linkcolor=blue,          % color of internal links, red is default
	citecolor=red,        % color of links to bibliography
	urlcolor=Violet            % color of external links, cyan is default
}
\usepackage[usenames,dvipsnames]{color}

\begin{document}

	\title{ 	
		\Large \bf{Dispersive approach to non-Abelian axial anomaly
%and \\ two-photon decays of pseudoscalar mesons
} }

\author{Sergey Khlebtsov,$^1$~~
	Yaroslav~Klopot,$^{2,3}$ ~~ 
	Armen~Oganesian,$^{1,2}$~~
	Oleg~Teryaev,$^2$
	\footnote{Electronic addresses: \href{mailto:klopot@theor.jinr.ru}{klopot@theor.jinr.ru}, \href{mailto:armen@itep.ru,}{armen@itep.ru,} \href{mailto:teryaev@theor.jinr.ru}{teryaev@theor.jinr.ru}.}
	\vspace{12pt} \\
	\it \small  $^1$Institute of Theoretical and Experimental Physics, 117218,  Moscow, Russia\\
	\it \small $^2$ %Bogoliubov Laboratory of Theoretical Physics,\\
	\it \small Joint Institute for Nuclear Research, 141980, Dubna, Russia\\
	\it \small $^3$Bogolyubov Institute for Theoretical Physics, 03143, Kiev, Ukraine}

\date{}
\maketitle
%\vfill

%	\textbf{Interplay between Abelian and non-Abelian anomalies in two-photon decays of mesons}\\

\textbf{Abstract.} 
Manifestations of strong and electromagnetic axial anomalies in two-photon decays of $\eta$ and $\eta'$ mesons are studied. Applying dispersive approach to axial anomaly in the singlet current, we obtain an anomaly sum rule containing strong and electromagnetic anomaly contributions. The relevant low energy theorem was generalized to the case of mixed states and used to evaluate the subtraction constant of the strong anomaly-related form factor  $\langle 0 |G\tilde{G} |\gamma\gamma \rangle$. We made a numerical estimation of the contributions of gluon and electromagnetic anomalies to the two-photon decays of $\eta$ and $\eta'$ mesons and found significant suppression of the gluon anomaly contribution.   \\

\textbf{1. Introduction.} \\

Axial (chiral) anomaly \cite{Adler:1969gk,Bell:1969ts} -- violation of the axial symmetry of classical theory by quantum fluctuations -- is an important phenomenon inherent to QCD  with many interesting consequences. In particular, axial anomaly is known to play an essential role in the two-photon decays of pseudoscalar mesons. As a matter of fact, it was the pion decay problem that had led to the discovery of quantum anomalies.  Precision measurements of two-photon decays of  $\pi^0$ \cite{Bernstein:2011bx, Ioffe:2007eg}, $\eta$ and $\eta'$  mesons remain a unique tool for the study of properties of QCD and effective theories at the low energy limit, including such subtle effects as chiral symmetry breaking and mixing.

Besides its connection with the real photon processes, the axial anomaly is intimately connected with the processes involving virtual photons: the dispersive form of the axial anomaly \cite{Dolgov:1971ri} (for a review, see e.g. \cite{Ioffe:2006ww}) leads to the anomaly sum rules (ASRs) \cite{Horejsi:1985qu,Horejsi:1994aj,Veretin:1994dn} which can be used to evaluate  the photon-meson transitions  $\gamma(k)\gamma^*(q)\to\pi^0(\eta, \eta')$ at arbitrary photon virtuality. This approach was used to study the transition form factors of the $\pi^0$, $\eta$ and $\eta'$ mesons in the space-like \cite{Klopot:2010ke,Klopot:2011qq,Klopot:2011ai,Melikhov:2012qp,Klopot:2012hd,Oganesian:2015ucv} and time-like \cite{Klopot:2013laa} regions. Along with the study based on the ASRs, the transition form factors  have been a subject of extensive investigation within other frameworks recently, such as light cone sum rules \cite{Agaev:2014wna,Stefanis:2012yw, Mikhailov:2016klg},  constituent \cite{Dorokhov:2013xpa}, light-front \cite{Choi:2017zxn} and non-local chiral  quark models \cite{GomezDumm:2016bxp}, light-front holographic QCD \cite{Brodsky:2011xx} as well as in some other models \cite{Roig:2014uja, Nedelko:2016vpj,Czyz:2017veo} and model-independent analyses \cite{Escribano:2013kba,Escribano:2015yup,Hanhart:2013vba}.

The presence of the axial anomaly results in the non-conservation of the axial current (even in the chiral limit). For the axial current $J_{\mu 5}=\bar{\psi_i}\gamma_\mu\gamma_5\psi_i$, where $\psi_i$ is some quark field of unit charge $e$, the axial anomaly leads to

\begin{equation}
\partial^{\mu} J_{\mu 5}=2im_i \bar{\psi_i} \gamma_5 \psi_i + \frac{e^2}{8\pi^2} N_c F\widetilde{F} + \frac{\alpha_s}{4\pi} G\widetilde{G},
\end{equation}
where $F$ and $G$ are  electromagnetic and gluon field strength tensors respectively,  $\widetilde{F}^{\mu\nu}=\frac{1}{2}\epsilon^{\mu\nu\rho\sigma}F_{\rho\sigma}$ and $\widetilde{G}^{\mu\nu,a}=\frac{1}{2}\epsilon^{\mu\nu\rho\sigma}G_{\rho\sigma}^{a}$ are their duals, $N_c=3$ is a number of colors, $\alpha_s$ is a strong coupling constant. 

In the case of light pseudoscalar mesons the relevant currents are the diagonal components of the octet of axial currents   $J_{\mu5}^{(a)}=(1/\sqrt{2})\sum_i {\bar{\psi_i}\gamma_{\mu}\gamma_5\lambda^a  \psi_i}$ and the singlet axial current $J_{\mu 5}^{(0)}=(1/\sqrt{3})\sum_i {\bar{\psi_i}\gamma_{\mu}\gamma_5 \psi_i}$,  where the sum is over the flavors of light quarks $i=u,d,s$,  $\lambda^a$ are the diagonal Gell-Mann $SU(3)$ matrices, $a=3,8$.  While the $\pi^0$ is almost pure $SU(3)$ flavor state (with corresponding $J_{\mu5}^{(3)}$ current), the  $\eta$ and $\eta'$ mesons are not -- their physical states are a significant mixture of the octet and singlet $SU(3)$ states (related to $J_{\mu5}^{(8)}$ and $J_{\mu5}^{(0)}$ currents). The mixing in the $\eta-\eta'$ system results in four non-zero decay constants $f^{(i)}_M$, defined as the currents' projections onto meson states $M$ ($i=8,0; M=\eta, \eta'$),
\begin{equation} \label{def_f}
\langle 0|J^{(i)}_{\mu 5}(0) |M(p)\rangle=
i p_\mu f^i_M.
\end{equation}

It is important, that the octet of axial currents is free from the strong (gluon) anomaly part while the singlet axial current acquires both electromagnetic as well as gluon anomalies,

\begin{equation}\label{dja}
\partial^\mu J_{\mu 5}^{(a)} =\frac{2i}{\sqrt2}\sum_i{m_i \bar{\psi_i} \gamma_5 \lambda^a \psi_i} 
+\frac{e^2}{8\pi^2}C^{(a)}N_c  F\widetilde{F}, \; a=3,8,
\end{equation}

\begin{equation}
\partial^\mu J_{\mu 5}^{(0)} =\frac{2i}{\sqrt 3}\sum_i{m_i \bar{\psi_i} \gamma_5 \psi_i}+\frac{e^2}{8\pi^2}C^{(0)}N_c  F\widetilde{F} + \frac{\sqrt{3}\alpha_s}{4\pi}  G\widetilde{G},
\end{equation}
where $C^{(a)}$ are the charge factors ($e_i$ are quark charges in units of the electron charge):
\begin{align}
C^{(3)}=&\frac{1}{\sqrt 2} (e_u^2-e_d^2)=\frac{1}{3\sqrt 2},\\ C^{(8)}=&\frac{1}{\sqrt 6} (e_u^2+e_d^2-2e_s^2)=\frac{1}{3\sqrt 6},\\
C^{(0)}=&\frac{1}{\sqrt 3} (e_u^2+e_d^2+e_s^2)=\frac{2}{3\sqrt 3}. 
\end{align} 
%For the 3rd (isovector) and the 8th (octet) components of the octet of axial currents, which are relevant for $\pi^0$,  $\eta$ and $\eta'$ 2-photon decays, the anomaly sum rules allowed to perform an extensive study of the meson transition form factors at arbitrary virtualities of the photon (processes $\gamma\gamma^*\to\pi^0(\eta,\eta')$) \cite{Klopot:2010ke,Klopot:2011qq,Klopot:2011ai,Klopot:2012hd,Melikhov:2012qp, Klopot:2013laa}.

Absence of the gluon anomaly for the  3rd (isovector) and the 8th (octet) components of the octet of axial currents allowed to derive the anomaly sum rules \cite{Klopot:2010ke,Klopot:2011qq,Klopot:2012hd,Melikhov:2012qp, Klopot:2013laa}  which benefited from the absence of corrections due to Adler-Bardeen theorem and t'Hooft's principle.  

The singlet axial current has a complication due to gluon anomaly part. This paper is aimed to investigate this issue.  We derive the anomaly sum rule based on the dispersive form of axial anomaly in the singlet channel and study the contributions of electromagnetic and gluon parts of the axial anomaly to the two-photon decays of the $\eta$ and $\eta'$ mesons.

The paper is organized as follows. In the Section 2 we derive the anomaly sum rule for the singlet axial current. In the Section 3 we generalize the low energy theorem for the case of mixing ($\eta-\eta'$) states. In the Section 4 we apply the results of the previous sections to study the role of electromagnetic and gluon parts of the axial anomaly in the meson decays.
\\

\textbf{2. Dispersive approach to axial anomaly with gluon term}\\  

In order to study the hadron observables in the non-perturbative region, we will develop a sum rule basing on the dispersive representation of axial anomaly in the singlet current. 
Consider the triangle graph amplitude, composed of the axial current $J_{\alpha 5}$ with momentum $p=k+q$ and two vector currents  with momenta $k$ and $q$ 
\begin{equation} \label{VVA}
\int
d^4 x d^4 y e^{(ikx+iqy)} \langle 0|T\{ J_{\alpha 5}(0) J_\mu (x)
J_\nu(y) \}|0\rangle=e^2T_{\alpha \mu\nu}(k,q). 
\end{equation}
This amplitude can be decomposed \cite{Rosenberg:1962pp} (see also \cite{Eletsky:1982py,Radyushkin:1996tb}) as

\begin{align}
\label{eq1} \nonumber T_{\alpha \mu \nu} (k,q)  & =  F_{1} \;
\varepsilon_{\alpha \mu \nu \rho} k^{\rho} + F_{2} \;
\varepsilon_{\alpha \mu \nu \rho} q^{\rho}
\\ \nonumber
& + \; \; F_{3} \; k_{\nu} \varepsilon_{\alpha \mu \rho \sigma}
k^{\rho} q^{\sigma} + F_{4} \; q_{\nu} \varepsilon_{\alpha \mu
	\rho \sigma} k^{\rho}
q^{\sigma}\\
& + \; \; F_{5} \; k_{\mu} \varepsilon_{\alpha \nu
	\rho \sigma} k^{\rho} q^{\sigma} + F_{6} \; q_{\mu}
\varepsilon_{\alpha \nu \rho \sigma} k^{\rho} q^{\sigma},
\end{align}
where the coefficients $F_{j} = F_{j}(p^{2},k^{2},q^{2}; m^{2})$, $j = 1, \dots ,6$ are the corresponding Lorentz invariant amplitudes constrained by current conservation and Bose symmetry. Note that the latter includes the interchange $\mu \leftrightarrow \nu$, $k \leftrightarrow q$ in the tensor structures and $k^2 \leftrightarrow q^2$ in the arguments of the scalar functions $F_{j}$.

Anomalous axial-vector Ward identity for $T_{\alpha \mu \nu} (k,q)$ for the singlet axial current $J_{\mu 5}^{(0)}(p)$ and photons $\gamma (k,\epsilon^{(k)})$, $\gamma(q, \epsilon^{(q)})$ (real or virtual) reads

\begin{equation}
p_{\alpha}T^{\alpha \mu \nu} = 2\sum_i m_iG_i\epsilon^{\mu \nu \rho \sigma} k_\rho q_\sigma +
\frac{C^{(0)} N_c}{2\pi^2}\epsilon^{\mu \nu \rho \sigma } k_\rho q_\sigma + N(p^2,q^2,k^2)\epsilon^{\mu
	\nu \rho \sigma} k_\rho q_\sigma,
\end{equation}
where the sum is over $i=u,d,s$ and 
\begin{equation}
\langle 0 | \frac{1}{\sqrt 3}\sum_{i} m_i\bar{\psi_i}\gamma_5\psi_i|\gamma\gamma\rangle =2\sum_i m_i G_i \epsilon^{\mu \nu \rho \sigma} k_\rho q_\sigma\epsilon_{\rho}^{(k)}\epsilon_{\sigma}^{(q)},
\end{equation}
\begin{equation} \label{N}
\langle 0 | \frac{\sqrt{3}\alpha_s}{4\pi} G\tilde{G}|\gamma\gamma \rangle = e^2 N(p^2, k^2,q^2) \epsilon^{\mu\nu\rho\sigma}k_{\mu}q_{\nu}\epsilon_{\rho}^{(k)}\epsilon_{\sigma}^{(q)},
\end{equation}

\begin{equation}
\langle 0 | F\tilde{F}|\gamma\gamma \rangle =  2\epsilon^{\mu\nu\rho\sigma}k_{\mu}q_{\nu}\epsilon_{\rho}^{(k)}\epsilon_{\sigma}^{(q)}.
\end{equation}
We introduced here the form factors $G_i$ and $N$, while the last matrix element is point-like up to QED corrections.

In the kinematical configuration with one real photon ($k^2=0$) which we consider in the rest of this section, the above anomalous Ward identity can be rewritten in terms of form factors $F_j,  G_i, N$ as follows ($N(p^2,q^2)\equiv N(p^2,q^2, k^2=0)$):
\begin{equation} \label{dr1}
(q^2 - p^2)F_3 - q^2F_4 = \sum_i 2m_iG_i + \frac{C^{(0)} N_c}{2\pi^2}+N(p^2,q^2).
\end{equation}
We can write the form factors $G_i, F_3, F_4$  as dispersive integrals without subtractions. Indeed, in the case of isovector and octet channels (free from gluon anomaly) it can be shown explicitly \cite{Horejsi:1994aj}. In the considered case of the singlet current from simple dimensional arguments one can assume that  $G_i, F_{3,4}$ decrease at large $p^2$, and therefore, the form factors can be written as dispersive integrals without subtractions. On the other hand, generally speaking, one should get the subtraction constant in the dispersion relation for the form factor $N$, analogous to the Abelian anomaly constant $\frac{C^{(0)} N_c}{2\pi^2}$. Let us rewrite this dispersion relation in the form with one subtraction at $p^2=0$:
\begin{equation}N(p^2,q^2)=N(0, q^2) + p^2 R (p^2,q^2),
\end{equation}
where the new form factor $R$ can be written as an unsubtracted dispersive integral.
Then the imaginary part of (\ref{dr1}) w.r.t. $p^2$ ($s$ in the complex plane) reads 
\begin{equation} \label{im}
(q^2 - s)\textit{Im}F_3 - q^2\textit{Im}F_4 = 2\sum_i m_i \textit{Im}G_i + s\textit{Im}R.
\end{equation}
Dividing every term of Eq. (\ref{im}) by $(s-p^2)$ and integrating over $s\in [0,+\infty)$, we get\footnote{The lower limits of the integrals are formally expressed in terms of quark masses, but due to confinement they should be replaced with a pion mass  (see, e.g., \cite{Gorsky:1989qd}), which we neglect anyway.}:

\begin{equation} \label{disp}
\frac{1}{\pi}\int_{0}^{\infty}\frac{(q^2-s)\textit{Im}F_3}{s-p^2}ds
- \frac{q^2}{\pi}\int_{0}^{\infty}\frac{\textit{Im}F_4}{s-p^2}ds =
\frac{1}{\pi}\sum_i \int_{0}^{\infty}\frac{2m_i\textit{Im}G_i}{s-p^2}ds +
\frac{1}{\pi}\int_{0}^{\infty}\frac{s\textit{Im}R}{s-p^2}ds.
\end{equation}

% Here, for simplicity, we put the limits to zero.
After simple transformation of the first and last terms in (\ref{disp}) and making use of the dispersive relations for the form factors $F_3, F_4, G_i, R$ we arrive at  

\begin{equation}\label{dr3}
(q^2-p^2) F_3 -\frac{1}{\pi}\int_{0}^{\infty} Im F_3 ds -q^2F_4 = 2\sum_im_iG_i + p^2R + \frac{1}{\pi}\int_{0}^{\infty} Im R ds.
\end{equation}
Comparing now (\ref{dr3}) with (\ref{dr1}) we can write down the anomaly sum rule for the singlet current:
\begin{equation} \label{asrfin}
\frac{1}{\pi}\int_{0}^{\infty}\textit{Im}F_3ds = \frac{C^{(0)} N_c}{2\pi^2} + N(0,q^2)
- \frac{1}{\pi}\int_{0}^{\infty}\textit{Im}R(s,q^2)ds,
\end{equation}
Saturating the l.h.s. of (\ref{asrfin}) with resonances according to global quark-hadron duality, we write out the first resonances' contributions explicitly,  while the higher states are absorbed by the integral with a lower limit $s_0$,

\begin{equation} \label{asr}
\sum_M {f_M^0F_{M\gamma} (q^2)}+ \frac{1}{\pi}\int_{s_0}^{\infty}\textit{Im}F_3ds =
\frac{C^{(0)} N_c}{2\pi^2} + N(0,q^2) -
\frac{1}{\pi}\int_{0}^{\infty}\textit{Im}R(s,q^2)ds,
\end{equation}
where the hadron contributions are expressed in terms of the decay constants $f_M^0$ (\ref{def_f}) and form factors $F_{M\gamma}(q^2)$ of the transitions $\gamma\gamma^* \to M$
\begin{equation}
\int d^{4}x e^{ikx} \langle M(p)|T\{J_\mu (x) J_\nu(0)
\}|0\rangle = e^2 \epsilon_{\mu\nu\rho\sigma}k^\rho q^\sigma
F_{M\gamma}(q^2) \;.
\end{equation}

The lower limit $s_0(q^2)$ in the integral in the l.h.s. of (\ref{asr})  should range between the masses squared of the last taken into account resonance and the first resonance included into the integral term. The choice of $s_0$ for the isovector and octet channels was discussed earlier \cite{Klopot:2011ai,Oganesian:2015ucv}. For the case of singlet current, keeping $\eta$ and $\eta'$ mesons in the first term of (\ref{asr}), we expect $s_0\gtrsim 1$~GeV$^2$. Actually, this estimation is sufficient for the purposes of the present paper.

As a note, let us point out at the following observation. We can also saturate with resonances the last term in the ASR (\ref{asrfin}). The main contributions are given, in particular, by the glueball-like states. Although it is hard to draw any numerical conclusions at present (for instance, the decay constants of the respective states are not known), the ASR can be useful for estimation of their relative contributions in the future.\\

\textbf{3. Low-energy theorem and mixing} \\

An important part of the ASR (\ref{asr}), representing gluon anomaly, is related to the matrix element $\langle 0 |G\tilde{G}(p)|\gamma(k)\gamma(q) \rangle$. 
Rigorous QCD calculation of this matrix element encounters difficulties due to confinement and is not known yet. However, it is possible to estimate it in the limit  $p^\mu=0$. Hereafter, we consider the case of two real photons ($q^2=k^2=0$).

The idea is simple (see \cite{Shifman:1988zk} and references therein). Consider the matrix element of the singlet axial current $ \langle 0 |   J_{\mu 5}^{(0)}(p)|\gamma\gamma\rangle$. Supposing that there are no massless particles in the singlet channel in the chiral limit, as the $\eta'$ meson remains massive, one must get $\lim\limits_{p\to 0} p^\mu \langle 0 |   J_{\mu 5}(p)|\gamma\gamma\rangle = 0$. This corresponds to $\langle 0 |   \partial^\mu J_{\mu 5}|\gamma\gamma\rangle = 0$, so using the explicit expression for the divergence of axial current in the chiral limit (put $m_q=0$), one can relate the matrix elements of $\langle 0| G\tilde{G}| \gamma\gamma \rangle$ and $\langle 0| F\tilde{F}|\gamma\gamma\rangle$  in the considered limits. 

However, due to a significant mixing in the $\eta-\eta'$ system,  the assumption of \cite{Shifman:1988zk}, that the singlet channel in the chiral limit does not contain massless particles, is violated by the contribution of the massless in the chiral limit $\eta$. Therefore, our aim now is to construct such a current, that has no projections onto the Goldstone states. Taking into account that $\pi^0$ meson has a negligible projection onto  $J_{\mu 5}^{(8)}$ and $J_{\mu 5}^{(0)}$ ($\sim$1\%  \cite{Gross:1979ur,Ioffe:1979rv}), we can limit our basis to these currents and require the current to be orthogonal only to $\eta$:

\begin{align}
J_{\mu 5}^{(X)}=a J_{\mu 5}^{(0)} + b J_{\mu 5}^{(8)}, \;\; \langle 0 |  J_{\mu 5}^{(X)}|\eta\rangle = 0.
\end{align}
After eliminating the constant $a$,  in terms of meson decay constants this current reads
\begin{equation}\label{jx}
J_{\mu 5}^{(X)}=b(J_{\mu 5}^{(8)} - \frac{f_{\eta}^8}{f_{\eta}^0} J_{\mu 5}^{(0)}),
\end{equation}
where b is an arbitrary constant and the decay constants $f_M^{(i)}$ are the defined in (\ref{def_f}).
The current (\ref{jx}) gives no massless poles in the matrix element $ \langle 0 |   J_{\mu 5}^{(x)}|\gamma\gamma\rangle$ even in the chiral limit, so
\begin{equation}
\lim\limits_{p\to 0} \langle 0 |   \partial_\mu J_{\mu 5}^{(X)}(p)|\gamma\gamma\rangle = 0.
\end{equation} 
Using explicit expressions for the divergences of currents in the chiral limit, at  $p^{\mu}=0$ we immediately obtain the following relation between the matrix elements of $G\tilde G$ and $F\tilde F$:

\begin{equation} \label{let}
\langle 0 | \frac{\sqrt{3} \alpha_s}{4\pi}G\tilde{G}|\gamma\gamma \rangle = \frac{N_c}{f_{\eta}^8}(f_\eta^{0} C^{(8)} -f_\eta^{8}C^{(0)})\langle 0 | \frac{\alpha_e}{2\pi} F\tilde{F}|\gamma\gamma \rangle.
\end{equation}
This gives us the value of the subtraction constant of the gluon anomaly,
\begin{equation} \label{n00}
N(0,0,0) = \frac{N_c}{2\pi^2 f_{\eta}^8}(f_\eta^{0} C^{(8)} -f_\eta^{8}C^{(0)}). 
\end{equation}

\textbf{4. Hadron contributions and analysis of the ASR} \\

As we mentioned above, the first hadron contributions to the ASR (\ref{asr}) are given by $\eta$ and $\eta'$. We keep these resonances as explicit contributions, while the rest of the resonances are absorbed by the integral "continuum" term.  In what follows, we limit ourselves to  the case of real photons, i.e. $k^2=q^2=0$. In this limit the transition form factors determine the two-photon decay amplitudes $A_M$ ($M=\eta, \eta'$) which are expressed in terms of the decay widths of the mesons $\Gamma_{M\to2\gamma}$ as follows:
\begin{equation}
A_M\equiv F_{M\gamma}(0)=\sqrt{\frac{64\pi\Gamma_{M\to2\gamma}}{e^4 m_M^3}}. \label{am}
\end{equation}
Recall also, that the ASR for the octet channel \cite{Klopot:2012hd} in the case of real photons leads to 
\begin{equation}
f_\eta^{8} A_{\eta} +f_{\eta'}^{8} A_{\eta'} = \frac{1}{2\pi^2}N_c C^{(8)}. \label{asr8}
\end{equation}

The ASR (\ref{asr}) for the singlet channel for real photons can be written as follows:

\begin{equation}
f_\eta^{0} A_{\eta} +f_{\eta'}^{0} A_{\eta'} = \frac{1}{2\pi^2} N_cC_0 + B_0 +B_1, \label{asr0}
\end{equation}
where, for the sake of brevity, we defined different contributions to the ASR as follows,
 
\begin{equation}
B_0\equiv N(0,0,0), \; B_1\equiv -\frac{1}{\pi}\int_{0}^{\infty}\textit{Im}R(s)ds -\frac{1}{\pi}\int_{s_0}^{\infty}\textit{Im}F_3ds.
\end{equation}
The $B_0$ term is the subtraction constant in the dispersion representation of gluon anomaly. The $B_1$ term consists of two parts: spectral representation of gluon anomaly and the integral covering higher resonances. The latter is proportional to $\alpha_s^2$. Indeed, the form factor $F_3$ is described by a triangle graph (no $\alpha_s$ corrections) plus diagrams with additional boxes ($\propto\alpha_s^2$ for the first box term). In the case of both real photons in the chiral limit the triangle amplitude is zero ($\propto q^2$). So, one can expect  $\alpha_s^2$ suppression of the higher resonances contributions term due to the sufficiently high lower limit of the integral, $s>s_0\gtrsim 1$~GeV$^2$.

Note, that at $s<s_0$ there is a NP QCD contribution following from (\ref{let}): $\langle 0 | G\tilde{G}|\gamma\gamma \rangle\propto\frac{\alpha_e}{\alpha_s}$. So, unlike the second term of $B_1$ (higher resonances contributions),  the first term of $B_1$ (spectral part of the gluon anomaly) lies in the essentially non-perturbative region.

Combining the ASRs for the octet (\ref{asr8}) and singlet (\ref{asr0}) channels of axial current, one gets:
\begin{equation} \label{aeta}
A_{\eta}= \frac{1}{\Delta}\left( \frac{N_c}{2\pi^2}(C^{(8)}f_{\eta'}^0-C^{(0)}f_{\eta'}^8)-(B_0+B_1)f_{\eta'}^8\right) , 
\end{equation}

\begin{equation} \label{aetap}
A_{\eta'}= \frac{1}{\Delta}\left( \frac{N_c}{2\pi^2}(C^{(0)}f_{\eta}^8-C^{(8)}f_{\eta}^0)+(B_0+B_1)f_{\eta}^8\right), 
\end{equation}
where $\Delta=f_{\eta}^{8}f_{\eta'}^{0}-f_{\eta'}^{8}f_{\eta}^{0}$.
Also, making use of the result of the low energy theorem (\ref{n00}) for $B_0$, we can express the two-photon decay amplitudes as follows,  
\begin{equation}
A_{\eta}= \frac{N_cC^{(8)}}{2\pi^2f_{\eta}^8}-\frac{B_1f_{\eta'}^8}{\Delta},
\end{equation}

\begin{equation}
A_{\eta'}= \frac{B_1f_{\eta}^8}{\Delta}.
\end{equation}
Note, that the low energy theorem leads to the cancellation of the photon anomaly term with subtraction part of gluon anomaly $B_0$ in (\ref{aetap}), so the  amplitude $\eta'\to~ \gamma\gamma$ (in the chiral limit) is entirely determined  by $B_1$, which is (predominantly) the spectral part of the gluon anomaly.

Let up pass to the numerical analysis. The $B_0 + B_1$ term can be evaluated  directly from the Eq. (\ref{asr0}) if we use the values of the two-photon decay widths and decay constants of the mesons. The low energy theorem additionally gives estimation for $B_0$, so combining (\ref{asr8}), (\ref{asr0}), we can separately evaluate $B_0$ and $B_1$.

\begin{table}
	\caption{Gluon anomaly term contributions for different sets of meson decay constants}
	\label{table1}
%	\begin{tabular}{ | l | c | l | l | l |}
\begin{tabular}{  l  c | l  l  l }
		\hline
		& 	$\left(\begin{array}{cc}  f_\eta^8 & f_{\eta'}^8 \\ f_\eta^0 & f_{\eta'}^0
		\end{array}\right)\frac{1}{f_\pi}$ & $B_0\times10^2$ & $B_1\times10^2$ & $(B_0+B_1)\times10^2$ \\ \hline
		\cite{Klopot:2012hd}, scheme-free  & $\left(\begin{array}{cc}  1.11 & -0.42 \\ 0.16 & 1.04 \end{array}\right)$ & -5.55 & 4.91 & -0.64 \\ %\hline
		\cite{Klopot:2012hd}, OS mix. scheme. & $\left(\begin{array}{cc}  0.85 & -0.22 \\ 0.20 & 0.81 \end{array}\right)$ & -5.36 & 3.84 & -1.53 \\ %\hline
		\cite{Klopot:2012hd}, QF mix. scheme  & $\left(\begin{array}{cc}  1.38 & -0.63 \\ 0.18 & 1.35 \end{array}\right)$ & -5.58 & 6.39 & 0.81 \\ %\hline
		\cite{Escribano:2005qq}, scheme-free & $\left(\begin{array}{cc}  1.39 & -0.59 \\ 0.054 & 1.29 \end{array}\right)$ & -5.77 & 5.86 & 0.095 \\ %\hline
		\cite{Feldmann:1998vh}, QF mix. scheme  & $\left(\begin{array}{cc}  1.17 & -0.46 \\ 0.19 & 1.15 \end{array}\right)$ & -5.51 & 5.47 & -0.047 \\ \hline
	\end{tabular}
\end{table}

For the decay constants $f_M^i$ we employ the sets of decay constants obtained in different analyses based on the octet-singlet (OS) mixing scheme \cite{Klopot:2012hd}, quark-flavor mixing scheme \cite{Klopot:2012hd,Feldmann:1998vh} and scheme-free approach  \cite{Klopot:2012hd,Escribano:2005qq}. The results are shown in the Table \ref{table1}.

These results demonstrate, that the contribution of the gluon anomaly and the  higher order resonances (expressed by $B_0+B_1$ term) to the 2-photon decay amplitudes appears to be rather small numerically in comparison with the contribution of electromagnetic anomaly $(1/2\pi^2)N_cC^{(0)}\simeq 0.058$. In fact, these processes are dominated by the electromagnetic anomaly: the electromagnetic part (the first two terms in (\ref{aeta}), (\ref{aetap})) makes $95\%$ and $90\%$ for $\eta$ and $\eta'$ meson decay amplitudes respectively, while the gluon anomaly originated part (the last two terms $\propto(B_0+B_1)$)  makes only $5\%$ and $10\%$ (for the decay constants scheme-free analysis from \cite{Klopot:2012hd}). Let us note, that this conclusion is valid for the processes with real photons: for the processes involving virtual photons (photon-meson transitions) it may not be true.

Using the low energy theorem gives the values of $B_0$ (subtraction constant) and, in combination with the results of the ASR (\ref{asr0}), $B_1$ (dominated by the term $\int_{0}^{\infty}\textit{Im}Rds$, the higher resonances term is suppressed as $\propto \alpha_s^2$, as we noted before). Numerically, $B_0$ and $B_1$  appear to be rather large: they are of order of the electromagnetic anomaly term.
At the same time, $B_0$ and $B_1$  enter the ASR with different signs and almost cancel each other, giving only a small total contribution to the two-photon decay widths of the $\eta$ and $\eta'$.
Our conclusions hold for different sets of decay constants which were obtained in independent analyses.\footnote{Somewhat different results for the constants of octet-singlet mixing scheme \cite{Klopot:2012hd} can be attributed to rather restricted properties of this scheme. Historically being the first one used for the $\eta-\eta'$ mixing description, nowadays it is rarely applied where precise analysis of the processes with $\eta-\eta'$ mixing is required.}% do not depend on the particular type of mixing scheme as it   \\
\\

\textbf{5. Conclusions and outlook} \\

Employing the dispersive approach to axial anomaly in the singlet current, we have obtained the sum rule with electromagnetic and gluon anomaly contributions. The gluon contribution consists of a spectral part (originated from $p^2$-dependent term) and a subtraction constant (independendent of $p^2$). 

The low energy theorem was generalized for the case of mixed $\eta-\eta'$ states and applied to evaluate the matrix element  $\langle 0 |G\tilde{G} |\gamma\gamma \rangle$ in the limit $p^\mu=0$. It gave an estimation for the subtraction constant of the gluon anomaly contribution in the dispersive form of axial anomaly.

The spectral part of the gluon anomaly was estimated using the ASR in the singlet current and low energy theorem result for the subtraction part. Numerically, it is found to be significant -- of the order of the electromagnetic anomaly contribution. However, it is almost canceled out by the subtraction term of gluon anomaly, resulting in the overall small contribution of the gluon anomaly to the $\eta(\eta')\to \gamma\gamma$ decays.

Also, application of the low energy theorem showed that the two-photon decay of $\eta'$ meson (in the chiral limit) is mainly determined by the spectral part of gluon anomaly. 

The smallness of the gluon contribution to radiative decays of pseudoscalar mesons may result in a relative suppression of the $\eta$ and $\eta'$ production from the color glass condensate in heavy ion collisions in favor of heavy glueballs. The properties of such glueballs may be deduced in a further analysis of the ASR  (\ref{asr}).

\textbf{Aknowledgments.}
We are thankful to A. Kataev,  M. Polyakov, N. Stefanis for useful discussions and illuminating comments. This work is supported in part by Heisenberg-Landau Program HL-2018 and by RFBR Grant 17-02-01108.

%\end{comment}
	
\end{document}